\documentstyle[12pt,thmsa,a4,sw20lart]{article}
\input tcilatex
\QQQ{Language}{
American English
}

\begin{document}

\title{Crystal-field interactions in PrRu$_2$Si$_2$. }
\author{R.~Michalski$^{a,b}$, Z. Ropka$^b$ and R.J. Radwa\'{n}ski$^{a,b}$ \\
%EndAName
$^a$Inst. of Physics, Pedagogical University, 30-084 Krak\'{o}w\\
$^b$Center for Solid State Physics, \'{s}w. Filip 5, 31-150 Krak\'{o}w\\
Poland.}
\maketitle

\begin{abstract}
Ferromagnetic compound PrRu$_2$Si$_2$ exhibits a giant magnetocrystalline
anisotropy of about 400 T. Its ordered moment below T$_c$=14K reaches 2.7 $%
\mu _B$ and is parallel to [0 0 1] crystalline direction. We have attributed
the magnetism of PrRu$_2$Si$_2$ to the Pr ions and performed calculations of
the fine electronic structure of the Pr ion in the tetragonal symmetry,
relevant to PrRu$_2$Si$_2$ taking into account crystal-field and inter-site,
spin-dependent exchange interactions. Spin-dependent interactions have been
taken into account by means of molecular-field approximation. The derived
energy level scheme is associated with the removal of the degeneracy of the
lowest multiplet given by Hund's rules, $^3$H$_4$. Magnetic and electronic
properties resulting from this fine structure are compared with all known
experimental results. Our calculations reproduce well the zero-temperature
moment, temperature dependence of the magnetic susceptibility,
single-crystalline magnetization curves with the anisotropy field of 400T,
the specific heat with the sharp peak at T$_c$ as well as
inelastic-neutron-scattering data.
\end{abstract}

\section{Introduction}

Ternary compounds RM$_2$X$_2$ (R= rare earth M= 3$d$, 4$d$ or 5$d$
transition metal and X=Si or Ge) exhibit large anisotropic magnetic
properties. PrRu$_2$Si$_2$ exhibits the largest measured magnetocrystalline
anisotropy. The anisotropy field at 4.2 K is estimated to be so enormous as
400 T \cite{1}. PrRu$_2$Si$_2$ crystallizes in the tetragonal structure ThCr$%
_2$Si$_2$-type and orders ferromagnetically below T$_C$=14K \cite{2}. Its
magnetic susceptibility shows strongly anisotropic behavior in temperatures
above T$_C$ \cite{3}. The temperature dependence of the specific heat of PrRu%
$_2$Si$_2$ shows the well-defined anomaly at the ferromagnetic ordering
temperature. The temperature dependence of the magnetic entropy computed
from the magnetic contribution shows value Rln2 at about 30K, indicating the
existence of two closely-lying levels. Inelastic-neutron-scattering
experiments \cite{1,4} reveal local excitations with energies of 2.25 and 29
meV providing strong argument for the existence of the localized CEF-like
levels. 

The aim of this paper is the evaluation, on basis of all known experimental
results, of the energy level scheme of the fine electronic structure of the
Pr$^{3+}$ ion in PrRu$_2$Si$_2$ associated with the configuration 4$f^2$.
This structure provides consistent description of main magnetic and
electronic properties of the compound under study, the direction and the
ordered moment value as well as the sharp peak at T$_C$ in the temperature
dependence of the heat capacity, in particular.

\section{Outline of theory}

In the individualized-electron model $f$ electrons keep their individuality
also being placed into intermetallic compounds. In the intermetallic
compound there coexists a few (here only 2 are specified) physically
important electronic subsystems: $f$ electronic subsystem(s) and
conduction-electron (c-e) subsystem. These two subsystems are described by
completely different theoretical approaches: localized-electron and
itinerant-electron models. The $f$ electron subsystem exhibits the discrete
energy spectrum associated with bound states of the electronic system $f^n$.
Itinerant electrons occupy the conduction band states. We have attributed
the magnetic properties of PrRu$_2$Si$_2$ to be predominantly due to the $%
4f^2$ electronic system of the Pr$^{3+}$ ions because the c-e susceptibility
is small and largely temperature independent as we know from the studies of
LaRu$_2$Si$_2$ \cite{5}. The Hund's rules ground multiplet is $^3$H$_4$ with 
$J$ = 4, $S$ = 1, $L$ = 5 and the Land\'{e} factor $g_L$ = 4/5. The general
Hamiltonian contains the single-ion-like and the intersite terms \cite{6,7}:

\begin{equation}
H=\sum \sum B_n^m\hat{O}_n^m(J,J_z)+ng_L^2\mu _B^2\left( -J\left\langle
J\right\rangle +\frac 12\left\langle J\right\rangle ^2\right) +g_L\mu
_BJ\cdot B_{ext}  \label{1}
\end{equation}

The first term is the CEF Hamiltonian written for the lowest multiplet given
by Hund's rules $^3$H$_4$. The second term represents the exchange
interactions between the Pr ions written in the mean-field approximation
with the molecular-field coefficient $n$. The third one describes the Zeeman
effect.

\section{Results and discussion}

The energy level scheme of the Pr ion in PrRu$_2$Si$_2$ for the tetragonal
symmetry CEF Hamiltonian contains 5 singlets and 2 doublets \cite{6}. A full
set of CEF parameters relevant to the tetragonal symmetry is given by: $B_2^0
$ = -22 K, $B_4^0$ = + 0.22 K, $B_4^4$ = + 0.20 K, $B_6^0$ = -12 mK and $%
B_6^4$ = -45 mK. This set of parameters has been derived by our
self-consistent analysis of experimental data. This set of CEF\ parameters
describes well

i) the energy separations revealed by the inelastic neutron scattering,

ii) anisotropic temperature dependence of the susceptibility in the
paramagnetic region,

iii) the direction of the magnetic moment (along the tetragonal c axis) in
the ordered state,

iv) the spontaneous moment of the Pr ion in 4.5 K of about 2.7 $\mu _B$,

v) the giant magnetocrystalline anisotropy of the magnetization curves at
4.5 K,

vi) temperature dependence of the specific heat with the $\lambda -$type
peak at T$_C.$

The consistent description of so many physical properties provides the
strong argument for our theoretical approach.

The ground state is a singlet in the form:

\begin{equation}
\Gamma _{t5}^{(1)}=0.703\left| +4\right\rangle +0.109\left| 0\right\rangle
+0.703\left| -4\right\rangle .  \label{2}
\end{equation}

The resulting energy-level scheme is shown in fig.1. It provides excitations
observable in inelastic-neutron-scattering (INS) experiments at energies of
30 K ($\Gamma _{t1}^{(1)}\rightarrow \Gamma _{t2}$) and 330 K ($\Gamma
_{t2}\rightarrow \Gamma _{t5}^{(1)}$ and $\Gamma _{t3}\rightarrow \Gamma
_{t5}^{(1)}$, $\Gamma _{t4}$). These excitations have been observed, indeed,
in INS experiments by Mulders et al. \cite{1,4} (2.25 and 29 meV). In this
experiment the 330 K excitation decreases with the increasing temperature.
This fact we take as the further confirmation of our scheme. With the
increasing temperature the contribution $\Gamma _{t3}\rightarrow $ $\Gamma
_{t4},$ having the lower energy, increases together with the Boltzmann
population of the $\Gamma _{t3}$ state. This observation provides indirect
argument for the existence of the state $\Gamma _{t3}^{(1)}$ at 58 K - the
neutron excitations to it from the lower states are prohibited. As will be
shown later the existence of a state at about 60 K is in very good agreement
with the overall temperature dependence of the specific heat. The entropy
analysis performed in Ref. 1 shows that there is a place for this extra
state owing to the increasing difference between the experimental and
calculated, for two states only, entropy  as one can see in Fig. 2 of Ref.1.

Closely lying first excited state together with the ground state creates the
interesting system. Their field-induced $J_z$ components have opposite signs
what lead to their opposite interactions with the magnetic field. As a
consequence this 2-levels system behaves as the quasi doublet. Moreover,
there is the large matrix element between these states. Thanks it the
magnetism of this system can be relatively easily induced by spin-dependent
interactions despite of general non-magnetic character of singlets. Such the
charge-formed ground state allows the appearance of the large magnetic
moment, of 2.7 $\mu _B,$ in agreement with experimental observation. Also
the direction of the moment is related with the shape of the eigenfunctions
of this 2-level system.

\subsection{Magnetic susceptibility}

In the quantum paramagnetic theory the magnetic moment is a property of the
electronic state of the paramagnetic ion. The value of the magnetic moment
of the ion reflects dependence of the energy of the ion with respect to the
magnetic field $B$ according to the definition \cite{8}: 
\begin{equation}
m(T)=-\frac{\delta E(T)}{\delta B}  \label{3}
\end{equation}
$E(T)$ is the total energy of the system over available electronic states
resulting from the Hamiltonian 1 and shown in Fig.1. The population of
states shown in Fig. 1 is given by the Boltzmann statistics. The magnetic
susceptibility in the paramagnetic region was calculated from the direct
diagonalization of the Hamiltonian, Eq. 1, without the second term.

The inverse of calculated temperature dependence of the magnetic
susceptibility in the paramagnetic region is shown in Fig.2. It exhibits
very anisotropic behavior with the easy c axis. The behavior $\chi ^{-1}(T)$
in the magnetic field parallel to the c axis almost follow the Curie-Weiss
law. The derived effective paramagnetic moment at room temperature amounts
to 4.18 $\mu _B$. It is larger than the free-ion value of 3.58 $\mu _B$ and
this fact is caused by strong CEF interactions. This paramagnetic moment
decreases with the increase of temperature approaching the theoretical value
at temperatures comparable with the overall size of the CEF\ interactions.
In the perpendicular field the calculated susceptibility is very small. The
visible discrepancy between calculated and experimental curves we attribute
to its small value that becomes comparable to the conduction-electron
contribution.

\subsection{Magnetic state and magnetocrystalline anisotropy}

At T$_c$ of 14 K magnetic order wins the temperature disordering. It is
accounted for in our calculations with the effective exchange interaction
parameter $n$ of 2.35 T/$\mu _B$. The magnetic ordering produces the abrupt
change in the energy level scheme that manifests in the $\lambda $ peak at T$%
_c$. The calculated moment of the Pr ion is parallel to the c axis in
agreement with experiment. It amounts to 2.7 $\mu _B$ at 0 K fully
reproducing the experimental datum \cite{1}. For completeness we add that it
turns out from our calculations that at 0 K the Pr$^{3+}$ ion in PrRu$_2$Si$%
_2$ experiences the internal molecular field of 6.4 T that originates from
the intersite RKKY interactions.

The magnetic moment of the Pr$^{3+}$ ion is strongly tied to the tetragonal
c axis as we see from the magnetization curves. In Fig. 3 the full
magnetization curves, at 4.5 K, calculated within our CEF approach are
presented. They are highly anisotropic: the field of 5.5 T applied along the
c-axis induces $\left\langle J_z\right\rangle $ of 3.6 what is about 100
times more then the value for $\left\langle J_x\right\rangle $. This
anisotropy preserves also in measurements up to 35 T, where magnetization
along the c and a axes is 3.08 $\mu _B$ and 0.39 $\mu _B$ , respectively 
\cite{1}. This large anisotropy is in agreement with single-crystal magnetic
measurements of Shigeoka et al.\cite{3}. The anisotropy field derived from
our calculated curves amounts to 400 T. This value corresponds to the
magnetocrystalline anisotropy energy of 59 J/cm$^3$, that is extremaly large.

For calculations of the magnetization curves in the ferromagnetic state
along the c-direction, shown in Fig. 3, the primary field-induced moment
curve $m$(B$_{full}$), shown in the insert, has been used. The extraction of
the influence of the external field $B_{ext}$, the internal field $%
B_{int}=n\cdot m$ has to be subtracted what leads to the metastable region
at zero field. Such the metastable situation means the spontaneous formation
of the ferromagnetic state.

The derived CEF parameters account also for the ferromagnetic ordering with
its direction, the ordering temperature and the size of the Pr-ion moment
and the giant magnetic anisotropy. These calculations reveal that the giant
anisotropy, even above 400 T, can be realized by the crystal-field
interactions.

\subsection{Specific heat}

The magnetic component of rare-earth specific heat is calculated by making
use of the general formula \cite{7}:

\begin{equation}
c_{4f}(T)=-T\frac{\delta ^2F(T)}{\delta T^2}  \label{4}
\end{equation}

$F(T)$ is the free energy of the rare-earth subsystem over the available
energy states resulting from the consideration of the Hamiltonian (eq. 1).
The calculated contribution of the Pr$^{3+}$ ions to the specific heat of
PrRu$_2$Si$_2$ is shown in fig.4. The overall behavior $c_{4f}(T)$ is in
very good agreement with experimentally derived magnetic contribution to the
specific heat \cite{1}. Our calculations reproduce well:

i) a $\lambda $-type peak at 14 K associated with the occurrence of the
ferromagnetic order. There is slight discrepancy in description of the shape
of the peak; the calculated one is wider then the experimental one. It may
indicate more dramatic character of the phase transition at T$_c$,

ii) a tail of the Schottky-like peak above T$_c$.

The reproduction of the $\lambda $ peak is important outcome of the present
paper. The previous calculations of Ref. 1 could not get the $\lambda $ peak
because they worked out with thermally independent electronic states. In
fact, the model of Ref.1 could not produce the ferromagnetic order.

This good agreement indicates on the substantial confidence to the presently
derived fine electronic structure. Our scheme gives much better description
of the overall $c_{4f}(T)$ dependence then previous calculations presented
in Ref. 1. They worked with the temperature independent structure of states
as they did not take into account the intersite spin-spin interactions and
they could not get the $\lambda $ peak at T$_c$.

\section{Conclusions}

The energy level scheme of the Pr$^{3+}$ ion in PrRu$_2$Si$_2$ has been
constructed on basis of all known experimental data. Our scheme is in
agreement with the inelastic-neutron-scattering data. Our calculations
reproduce well the zero-temperature moment and its temperature dependence,
temperature dependence of the magnetic susceptibility and the specific heat
with the sharp peak at T$_c$, single-crystalline magnetization curves with
the enormous anisotropy field of 400 T. The obtained consistent description
indicates on the substantial confidence to the present evaluation of the
crystal field interactions in PrRu$_2$Si$_2$ and should be the good starting
point for the CEF\ analysis of other members of RERu$_2$Si$_2$ class of
compounds.

{\bf Fig. 1.} The energy level scheme of the Pr$^{3+}$ ion in PrRu$_2$Si$_2$
with the expectation values of $J_z$ and eigenvectors in the paramagnetic
state. The tetragonal CEF interactions split the 9-fold degenerate ground
multiplet $^3$H$_4$ into 5 singlets and 2 doublets. Three arrows indicate
the allowed INS excitations. They are assigned to those experimentally
detected in Refs 1 and 4.

{\bf Fig. 2.} Temperature dependence of the magnetic susceptibility of PrRu$%
_2$Si$_2$ along the tetragonal c- and a-axis shown in the $\chi ^{-1}$ vs T
plot in the paramagnetic state. The solid lines show our calculations,
points are experimental data after Ref. 3.

{\bf Fig. 3.} Magnetization curves, the moment vs external field B$_{ext}$,
at 4.5 K (lines) on single-crystalline PrRu$_2$Si$_2$ calculated along main
crystallographic directions of the tetragonal unit cell in the ferromagnetic
state. Points denote experimental data from Ref. 1. In the insert the
primary field-induced magnetic moment is shown. In the ferromagnetic state,
for the extraction of the influence of the external field, the internal
field $B_{int}=n\cdot m$ is subtracted.

{\bf Fig. 4. }Temperature dependence of the 4$f$ specific heat contribution
in PrRu$_2$Si$_2$. The dashed-point line shows the result of our
calculations. The solid line represents calculations of Ref. 1. Points are
experimental data from Ref. 1.

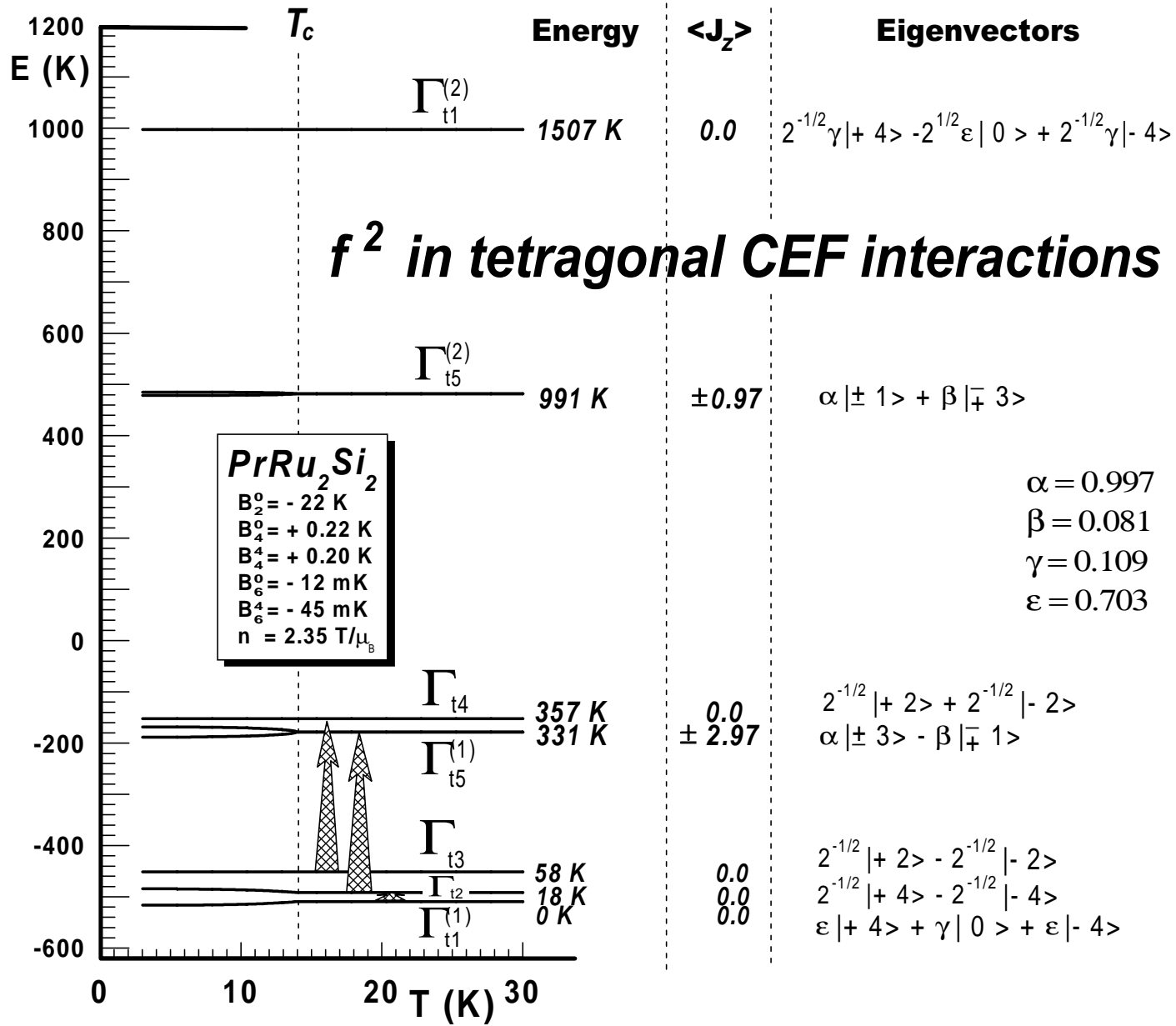
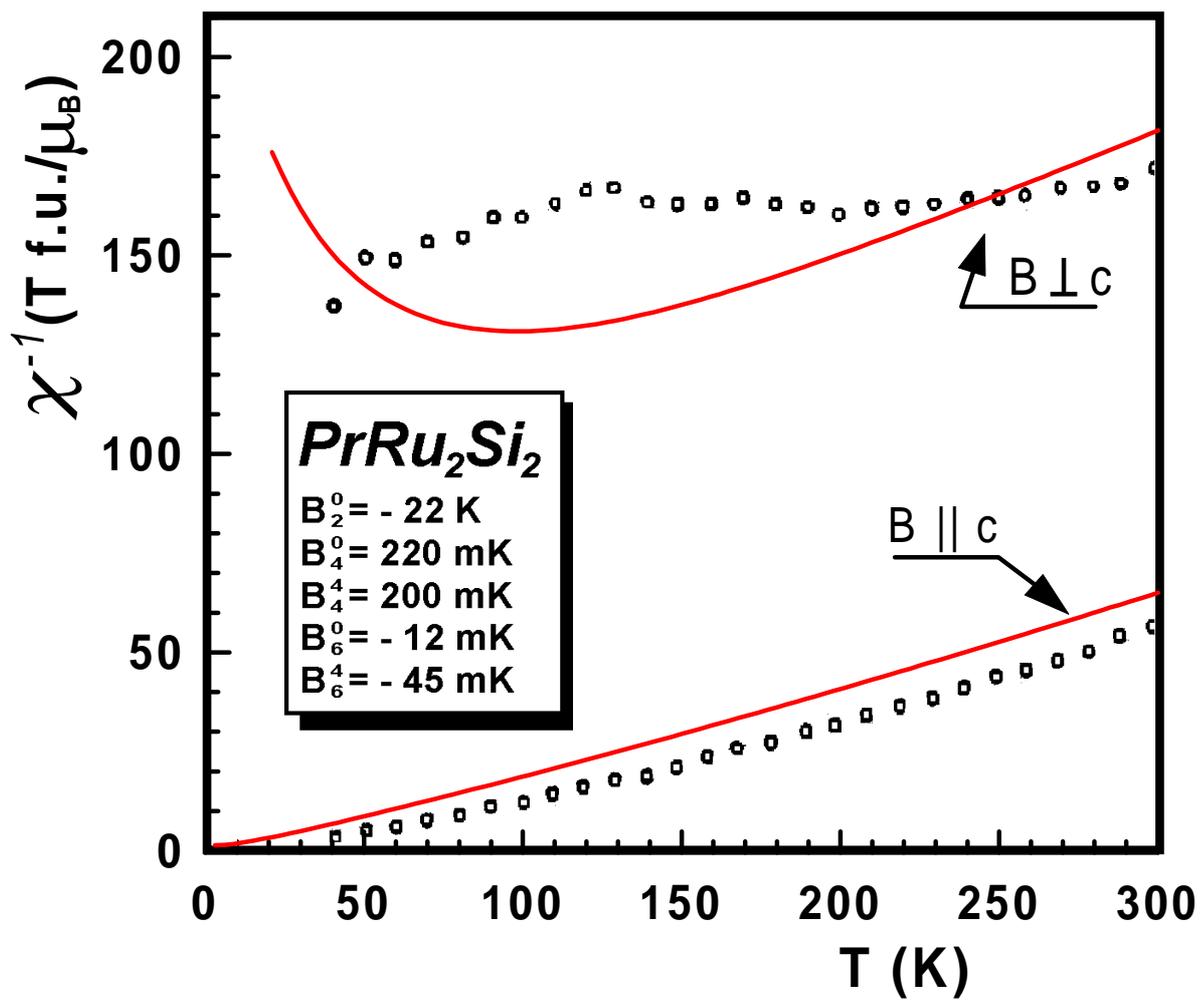
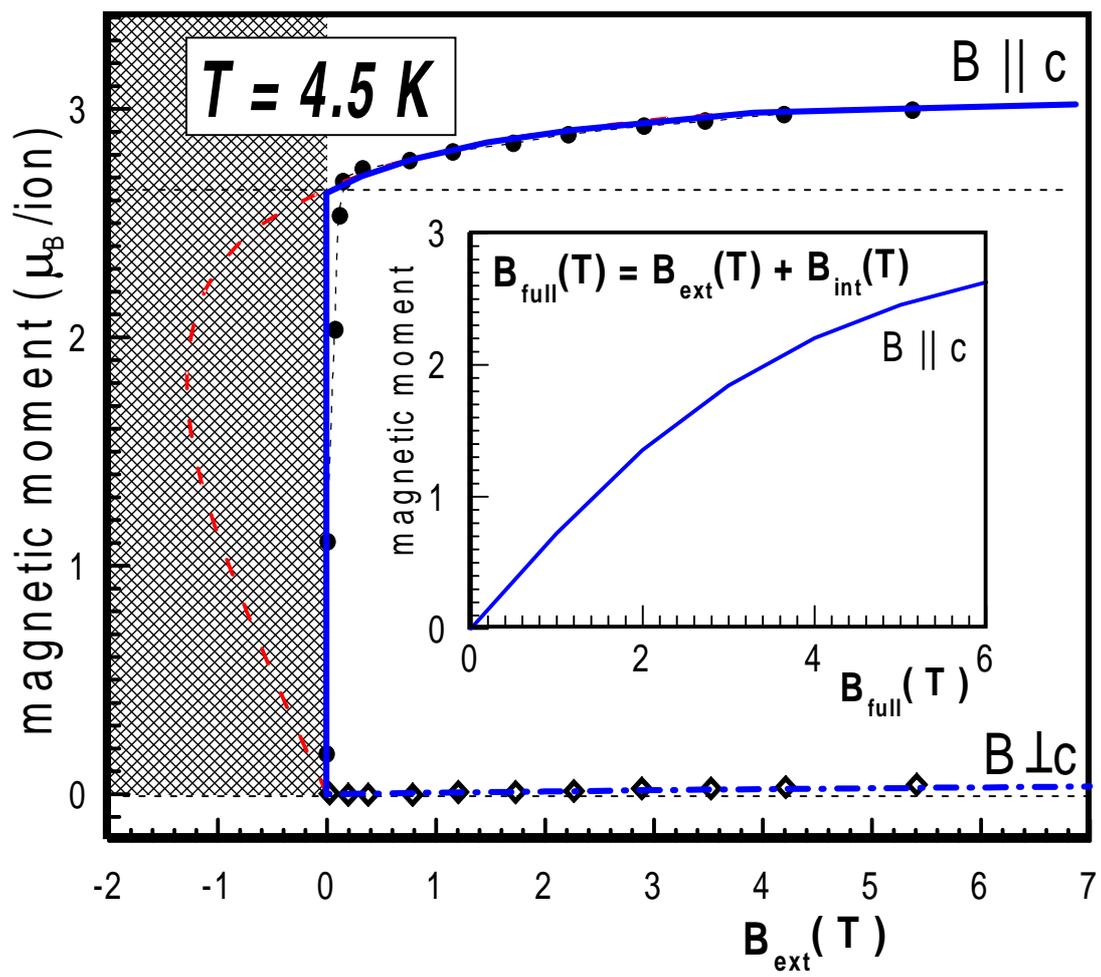
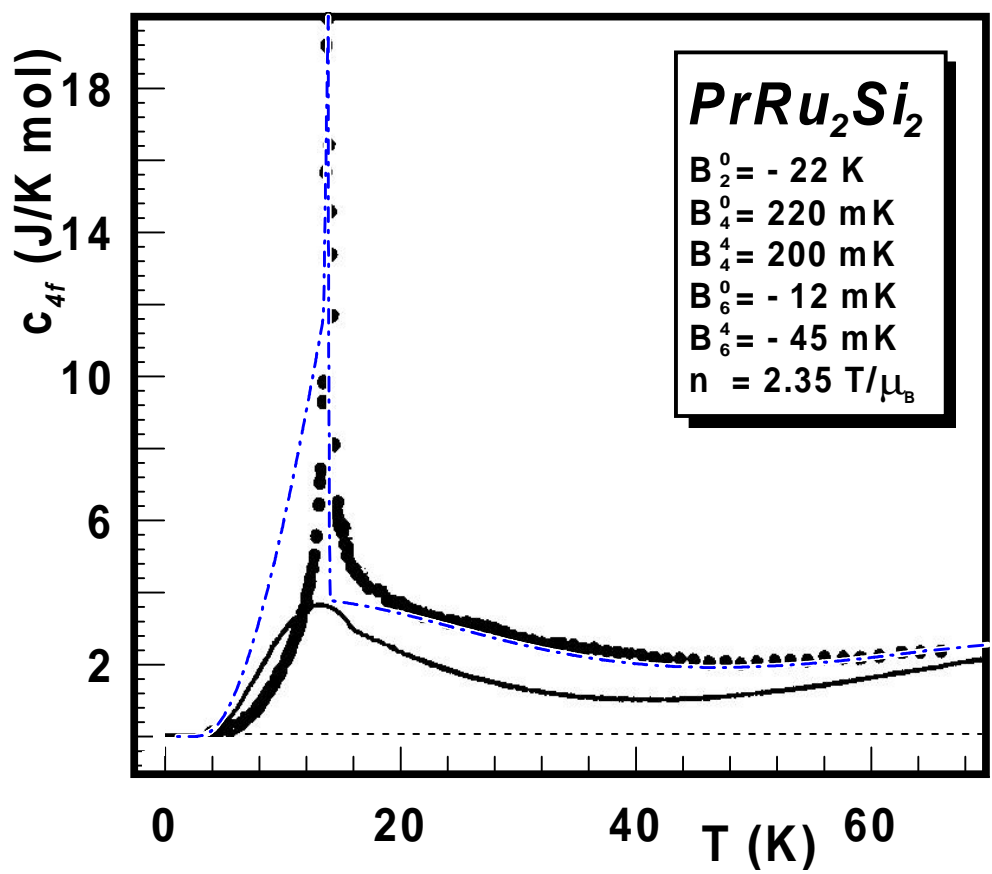
\end{document}